\documentclass[12pt,onecolumn,amsmath,amssymb, spacing=1.5pt]{article}
\usepackage[letterpaper,margin=1in]{geometry}
\usepackage{ulem}

\newcommand{\br}{\boldsymbol{\textbf{r}}}

\newcommand{\dr}{\,d\br}

\newcommand{\vxc}{v_{\rm{xc}}}
\newcommand{\rhoCI}{\rho_{\scriptsize{\text{CI}}}}
\newcommand{\rhoKS}{\rho_{\scriptsize{\text{KS}}}}

\newcommand{\Exc}{E_{\rm{xc}}}
\newcommand{\exc}{e_{\rm{xc}}}
\newcommand{\ExcNN}{E_{\rm{xc}}^{\rm{NN}}}
\newcommand{\excNN}{e_{\rm{xc}}^{\rm{NN}}}
\newcommand{\vxcNN}{v_{\rm{xc}}^{\rm{NN}}}

\usepackage{doi}
\usepackage{mathtools}
\usepackage{hyperref}
\hypersetup{
    colorlinks=true,
    linkcolor=magenta,
    filecolor=magenta,      
    urlcolor=cyan,
    }

\usepackage{subfiles}
\usepackage{graphicx}
\usepackage{graphics}
\usepackage{multirow}
\usepackage{amsxtra}
\usepackage{stmaryrd}
\usepackage{mathrsfs}
\usepackage{caption}
\usepackage{subcaption}
\usepackage{epsfig}
\usepackage{rotating}
\usepackage{setspace}
\usepackage{float}
\usepackage{amsfonts}
\usepackage{amsbsy}
\usepackage{amscd}
\usepackage{amsthm}
\usepackage{relsize}
\usepackage{color}
\usepackage{tablefootnote}
\usepackage{tabularx}
\usepackage{caption}
\usepackage{sidecap}
\usepackage{dcolumn}
\usepackage{footnote}
\usepackage{algorithm}
\usepackage{algorithmic}
\usepackage{breqn}
\usepackage{authblk}

\definecolor{hellgruen}{rgb}{0.2,0.7,0.2}
\newcommand{\cb}{\color{blue}}

\newcommand{\cn}{\color{black}}

\newcommand{\tcb}{\textcolor{blue}}
\newcolumntype{M}[1]{>{\centering\arraybackslash}m{#1}}
\newcolumntype{N}{@{}m{0pt}@{}}
\begin{document}
\title{Learning local and semi-local density functionals from exact exchange-correlation potentials and energies}
\author[a]{Bikash Kanungo}
\author[b]{Jeffrey Hatch}
\author[b]{Paul M. Zimmerman}
\author[a,c]{Vikram Gavini\footnote{Corresponding author. Email: vikramg@umich.edu}}
\affil[a]{\small Department of Mechanical Engineering, University of Michigan, Ann Arbor, Michigan 48109, USA}
\affil[b]{\small Department of Chemistry, University of Michigan, Ann Arbor, MI 48109, USA}
\affil[c]{\small Department of Materials Science and Engineering, University of Michigan, Ann Arbor, Michigan 48109, USA}
\date{}

\maketitle
\begin{abstract}
Finding accurate exchange-correlation (XC) functionals remains the defining challenge in density functional theory (DFT). Despite 40 years of active development, the desired chemical accuracy is still elusive with existing functionals. We present a data-driven pathway to learn the XC functionals by utilizing the exact density, XC energy, and XC potential. While the exact densities are obtained from accurate configuration interaction (CI), the exact XC energies and XC potentials are obtained via inverse DFT calculations on the CI densities. We demonstrate how simple neural network (NN) based local density approximation (LDA) and generalized gradient approximation (GGA), trained on just five atoms and two molecules, provide remarkable improvement in total energies, densities, atomization energies, and barrier heights for hundreds of molecules outside the training set. Particularly, the NN-based GGA functional attains similar accuracy as the higher rung SCAN meta-GGA, highlighting the promise of using the XC potential in modeling XC functionals. We expect this approach to pave the way for systematic learning of increasingly accurate and sophisticated XC functionals.          
\end{abstract}

\section{Introduction} \label{sec:intro}
For decades, density functional theory (DFT)~\cite{Hohenberg1964, Kohn1965, Becke2014} has been the workhorse of computational chemistry and materials science. Its ingenuity lies in formally reducing the
many-electron Schr\"odinger equation to an effective single electron problem governed by the electron density ($\rho(\br)$). Despite the formal exactness of the theory, the unavailability of the exact exchange-correlation functional necessitates the use of approximations. After 40 years of efforts at improving the XC functional \cb ~\cite{Cohen2012, Burke2012, Becke2014}\cn, XC approximations remain far from the desired quantum accuracy of 1 kcal/mol. This lack of accuracy among existing
XC approximations severely reduces quantitative reliability, limiting DFT to predicting qualitative trends in
material properties. Traditionally, XC approximations have been developed based on satisfying known conditions and/or
semi-empirical fitting, which, while useful, may not lend itself to systematic improvement. 

\noindent The resurgence and gradual percolation of machine-learning (ML) into science and engineering offers great promise for designing XC functionals, particularly with accurate training data. While a few attempts~\cite{Nagai2018, Zhou2019, Schmidt2019, Nagai2020, Dick2020, Kirkpatrick2021, Nagai2022} have explored the idea of using ML to model the XC functional, they present limitations in terms of their transferability, ability for systematic improvement, and adherence to crucial physical constraints. For instance, Refs.~\cite{Schmidt2019, Dick2020} are either based on model 1D systems or are developed to work within specific datasets, curtailing their transferability. Some past efforts~\cite{Nagai2018, Zhou2019} model the XC potential without any information of the XC energy, and hence, provide no guarantee of the potential's integrability. To elaborate, the modeled XC potential may not have a corresponding XC energy functional and hence can lead to different energies for different paths chosen to evaluate the XC energy via any path integral formulation---a severe violation of the fundamental aspect of the XC functional. The works of Nagai et.\ al.~\cite{Nagai2020, Nagai2022} present a more robust and theoretically well-grounded approach to modeling the XC functional by using the error in the predicted density, which generalizes well even when trained with limited data. However, this entails a self-consistent field iteration for each of the training systems at each epoch of machine-learning. As a result, this approach can become prohibitively expensive with increasing expressivity of the models and training samples. The DM21 functional~\cite{Kirkpatrick2021} is the most sophisticated ML-based XC model developed so far wherein a combination of three existing XC approximations are used with a learnable local hybrid enhancement factor for each of them. However, it is trained using a large dataset of 1141 molecules along with explicit use of a wide variety of thermochemical properties against which it is tested. This reliance on large dataset in DM21 can make it prone to poor generalization for systems and properties that are qualitatively different from those used in training\tcb{~\cite{Palos2022, Zhao2024}}. 

\begin{figure}
    \centering
    \includegraphics[scale=0.6]{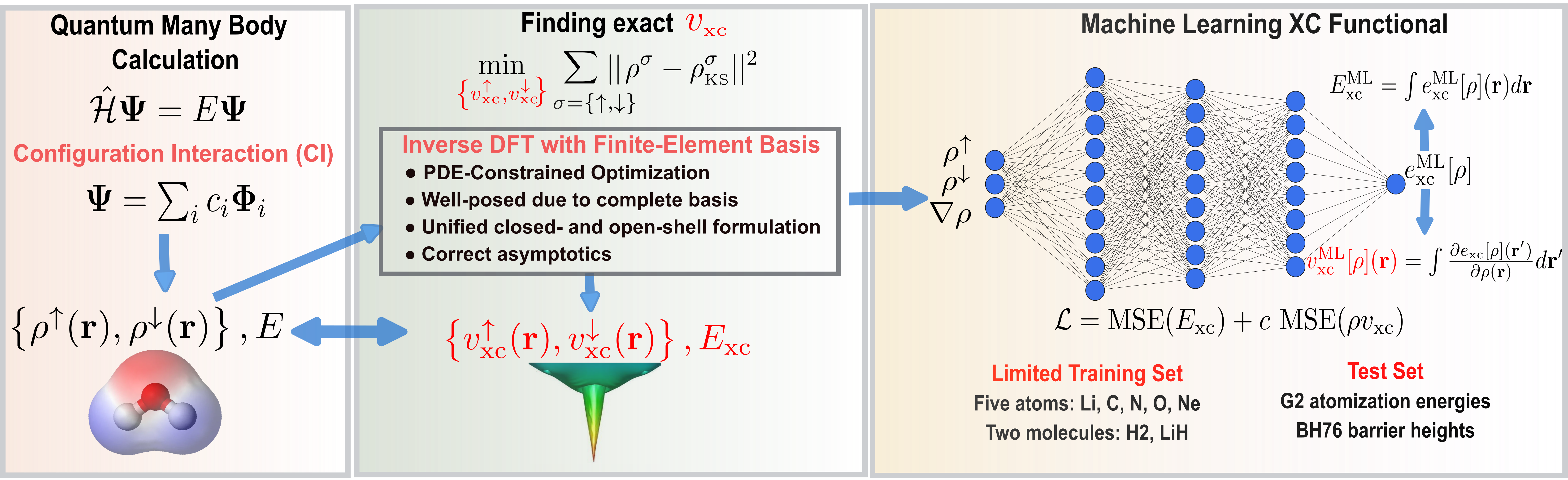}
    \caption{\small{Schematic of the ML based XC modeling. The exact spin-densities $\{\rho^{\uparrow}(\br),\rho^{\downarrow}(\br)\}$ and energies ($E$) are obtained from configuration interaction (CI) calculations. The inverse DFT with finite-element basis solves a PDE-constrained optimization to find the corresponding exact XC potentials $\{\vxc^{\uparrow}(\br),\vxc^{\downarrow}(\br)\}$  and the XC energy ($\Exc$). The neural network (NN) uses the exact XC densities, potentials, and energies to model the XC functional.}}  
    \label{fig:schematic}
\end{figure}

\noindent This work presents a pathway to learning transferable and data-efficient functionals by use of exact XC potentials ($\vxc(\br)$) in addition to XC energies ($\Exc$) (see schematic in Fig.~\ref{fig:schematic}). The use of the XC potential, as opposed to the typical use of density~\cite{Nagai2020, Nagai2022} and/or energy (or relative energy)~\cite{Dick2020, Kirkpatrick2021}, is motivated by three factors. First, the well-known bijective (one-to-one) map between the density and the XC potential promises greater generalizability with much less data. In contrast, the lack of any bijective relation between the density and the energy renders energy based models to suffer from the need of large training data and/or poor transferability. Second, compared to the density, the XC potential offers a more instructive quantity to be used as a target in any learning, owing to the fact that a small changes in the density can lead to large changes in the corresponding potential~\cite{Kanungo2021, Kanungo2023}. In other words, it is easier to distinguish two systems by their XC potential than by their densities. Third, the use of XC potentials bypasses the need of any self-consistent field calculation to measure the quality of the density during training---a costly proposition for complex ML-models and/or larger training data. 

\noindent We remark that although XC potentials have been used in prior ML-based XC modeling, they were either limited to model 1D systems~\cite{Schmidt2019} and/or provided no guarantee of integrability of the potential ~\cite{Nagai2018, Zhou2019}. More importantly, until recently, a robust and accurate solution to the \textit{inverse} DFT problem of obtaining the exact XC potential corresponding to a given density was challenging, owing to its numerous numerical difficulties~\cite{Shi2021}. As a result, the use of XC potentials to model the XC functional has seen limited progress. To that end, we employ our recent advances~\cite{Kanungo2019, Kanungo2023} in inverse DFT to find the exact XC potentials for various atoms and molecules from highly accurate ground state densities from configuration interaction (CI) calculations. 

\noindent The potency of the XC potential in modeling the XC functional will be demonstrated by learning neural-network (NN) based local and semi-local XC approximations, termed NNLDA and NNGGA, respectively. These approximations are analogous to the local density approximation (LDA) and generalized gradient approximation (GGA) in DFT. We use a compact training data comprising the exact density, XC potential, and XC energy of five atoms (Li, C, N, O, Ne), two dimers (H$_2$, LiH), and the uniform electron gas (UEG). Both NNLDA and NNGGA offer remarkable improvement in total energies, outperforming not just widely used LDA (PW92~\cite{Perdew1992}) and GGA (PBE~\cite{Perdew1996}) functionals but also B3LYP (a higher rung, hybrid functional). The NNGGA, in particular, attains a mean absolute error (MAE) of $<2$ kcal/mol in the total energy per atom, offering the best total energies among the functionals considered in this work. The NNGGA also significantly improves over PBE in the quality of its self-consistently solved density, as measured through the mean-field error~\cite{Gould2023b} (i.e., error in the non-XC part of their energies). Most importantly, the NNGGA functional attains nearly the same accuracy as the widely used SCAN~\cite{Sun2015} meta-GGA functional (a higher rung functional containing the additional information of the kinetic energy density) in both atomization energies and reaction barriers. The improvements in atomization energies and reaction barriers are noteworthy given that the training of the NN-based functionals did not use any of those quantities, unlike prior ML-based  efforts in XC modeling which directly used the target properties in their training~\cite{Nagai2020, Kirkpatrick2021, Nagai2022}. Overall, the promising accuracy of
NNGGA, trained with limited data, underscores the potency of using exact XC potentials in learning XC functionals.

\section{Results}\label{sec:results}

\subsection{Exact XC potentials and energies}
We briefly discuss our inverse DFT approach~\cite{Kanungo2019} that is pivotal to generating the exact XC potential ($\vxc(\br)$) and energy ($\Exc$) to be used as training data to model the XC functionals. 
Given an accurate ground state spin-density $\{\rho^{\uparrow}(\br), \rho^{\downarrow}(\br)\}$ (herein, from accurate CI calculations), finding the corresponding exact XC potentials ($\{\vxc^{\uparrow}(\br), \vxc^{\downarrow}(\br)\}$) can posed as the variational problem
\begin{equation} \label{eq:rhomin}
\text{arg}\min_{\{\vxc^{\uparrow}(\br), \vxc^{\downarrow}(\br)\}}\sum_{\sigma={\uparrow,\downarrow}}\int{w(\br)\left(\rho^{\sigma}(\br)-\rhoKS^{\sigma}(\br)\right)^2\dr}\,,
\end{equation}
subject to:
\begin{equation}\label{eq:KS} 
    \left(-\frac{1}{2}\nabla^2+v_{\text{ext}}(\br)+v_{\text{H}}(\br)+\vxc^{\sigma}(\br)\right)\psi^{\sigma}_{i}(\br)=\epsilon^{\sigma}_i\psi^{\sigma}_i(\br)\,,
\end{equation}
\begin{equation} \label{eq:Normal}
\int{\left(\psi^{\sigma}_i(\br)\right)^{*}\psi^{\sigma}_j(\br)\dr}=\delta_{i,j}\,.
\end{equation}
In the above equation, $v_{\text{ext}}$ and $v_{\text{H}}$ are the nuclear potential and Hartree potential (corresponding to the total density $\rho=\rho^{\uparrow}+\rho^{\downarrow}$), $\psi^{\sigma}_i$ and $\epsilon^{\sigma}_i$ are the Kohn-Sham eigenpairs for $\sigma$ spin; and $w(\br)$ is an appropriately chosen positive weight to expedite convergence, especially in the low density region. The Kohn-Sham density $\rhoKS^{\sigma}(\br)=\sum_{i}f^{\sigma}_i|\psi^{\sigma}_i(\br)|^2$, where $f^{\sigma}_i$ is the occupancy for $\psi^{\sigma}_i$. We solve the above constrained minimization problem using an adjoint state approach~\cite{Jensen2018, Kanungo2019}. While inverse DFT is susceptible to numerical artifacts in $\vxc^{\sigma}(\br)$, these artifacts have largely arisen from use of an incomplete basis (e.g., Gaussian basis) when discretizing the inverse problem~\cite{Jacob2011, Burgess2007, Bulat2007}.  Spurious oscillations can also be caused by the lack of correct asymptotics in the Gaussian densities obtained from CI, which lack a Kato-like cusp at the nuclei. Also, far from the nuclei, densities should also decay as exponentials, not Gaussians. These incorrect asymptotics in the Gaussian densities induce spurious oscillations in the resulting potential~\cite{Mura1997, Schipper1997, Gaiduk2013}. Our approach addresses these issues by using: (i) a systematically convergent finite-element basis, which renders the discrete problem well-posed; (ii) a small correction to the Gaussian densities to ensure cusps at the nuclei; and (iii) appropriate far-field boundary conditions on the XC potential to ensure the expected $-1/r$ decay. We refer to ~\cite{Kanungo2019, Kanungo2023} for details of the method, which reports on the accuracy and efficacy of our approach in obtaining accurate potentials. 

\noindent Having the solved the above inverse DFT problem, the exact XC energy ($\Exc$) can be evaluated from the exact ground state energy ($E$) and density $\rho(\br)=\rho^{\uparrow}(\br) + \rho^{\downarrow}(\br)$, computed from CI, as
\begin{equation} \label{eq:Exc}
    \Exc = E - \sum_{\sigma={\uparrow,\downarrow}}T_s[\rho^{\sigma}] -E_{\text{H}}[\rho] - \int\rho(\br)v_{\text{ext}}(\br)\dr\,,  
\end{equation}
where $T_s[\rho^{\sigma}]=\frac{1}{2}\sum_{i}f^{\sigma}_{i}\int|\nabla\psi^{\sigma}_i(\br)|^2\dr$ is the non-interacting kinetic energy obtained from the Kohn-Sham orbitals corresponding to the exact XC potential and $E_{\text{H}}[\rho$] denotes the Hartree energy of the density $\rho(\br)$. 

\subsection{Neural-network (NN) based functionals}
As is the standard practice in DFT, we define our NN-based XC energy functional ($\ExcNN[\rho]$) in terms of its energy density ($\excNN[\rho]$), defined as $\ExcNN[\rho]=\int\excNN[\rho](\br)\dr$.  In the spirit of local density approximation and generalized gradient appoximation (GGA), we model NN-based models that are analogously named as NNLDA and NNGGA, and are given by 
\begin{equation} \label{eq:NNLDA}
        e_{\text{xc}}^{\text{NNLDA}}[\rho](\br) = e_{\text{xc}}^{\text{PW92}}[\rho](\br) + e_{\text{x}}^{\text{UEG}}[\rho](\br)\phi(\br)G^{\text{NNLDA}}(\rho(\br),\xi(\br))\,.
\end{equation}
\begin{equation} \label{eq:NNGGA}
e_{\text{xc}}^{\text{NNGGA}}[\rho](\br) = e_{\text{xc}}^{\text{PBE}}[\rho](\br) + e_{\text{x}}^{\text{UEG}}[\rho](\br)\phi(\br)G^{\text{NNGGA}}(\rho(\br),\xi(\br),s(\br))\,.
\end{equation}
In the above, $e_{\text{xc}}^{\text{PW92}}[\rho]$ and $e_{\text{xc}}^{\text{PBE}}[\rho]$ are the PW92~\cite{Perdew1992} and PBE~\cite{Perdew1996} XC energy densities, used as base XC functionals. $e_{\text{x}}^{\text{UEG}}[\rho](\br)=-\frac{3}{4}\left(\frac{3}{\pi}\right)^{1/3}\rho^{4/3}(\br)$ is the exchange energy density of the uniform electron gas (UEG); $\xi$ is the relative spin density $\xi(\br)=(\rho^{\uparrow}(\br)-\rho^{\downarrow}(\br))/\rho(\br)$; $\phi(\br)=\frac{1}{2}\left((1+\xi(\br))^{4/3}+(1-\xi(\br))^{4/3}\right)$; and $s(\br)=|\nabla \rho(\br)|/[2(3\pi^2)^{1/3}\rho^{4/3}(\br)]$ is the reduced density gradient. The $G^{\text{NNLDA}}$ and $G^{\text{NNGGA}}$ are modeled as a neural network (NN) with $\{\rho, \xi\}$ and  $\{\rho, \xi, s\}$ as input descriptors, respectively.  
The explicit inclusion of spin information via $\xi$ is crucial in describing spin-polarized systems. The choice of the $\rho^{4/3}$ and $\phi$ as prefactors are chosen to nudge the model towards retaining known coordinate- and spin-scaling relations for the dominant exchange part of the functional. The XC energy density $\exc[\rho]$ is not uniquely defined, as one can add any functional of $\rho$ that integrates to zero without altering $\Exc$. Thus, there exists no exact $\exc[\rho](\br)$ for a system that can be used as a target during training of the models. We, therefore, use the XC potentials as our target to train NNLDA and NNGGA models. In particular, we use a composite loss function composed of mean squared errors (MSE) in XC energy ($\Exc$) and density-weighted XC potential ($\rho\vxc$), given as
\begin{equation} \label{eq:}
    \mathcal{L} = \frac{c}{M} \sum_{I=1}^M \left(E_{\text{xc},I} - \int \excNN[\rho_{I}](\br)\dr\right)^2 + \frac{1}{M} \sum_{I=1}^M \sum_{\sigma={\uparrow,\downarrow}}\left(\int \rho^{\sigma}_{I}(\br)\left(v_{\text{xc},I}^{\sigma}(\br)-v_{\text{xc}}^{\sigma,\text{NN}}[\rho_I](\br)\right)\dr\right)^2\,, 
\end{equation}
where $I$ indexes the training samples and the constant $c$ is used to weight the energy and potential loss terms differently. The quantities without the NN superscript refer to the exact values obtained from inverse DFT calculations on accurate CI densities. The $\vxcNN$, for a local/semi-local functional, is given as $v_{\text{xc}}^{\sigma,\text{NN}}[\rho](\br)=\frac{\delta \excNN[\rho](\br)}{\delta \rho^{\sigma}(\br)}$ and can be inexpensively evaluated using back-propagation. The use of density-weighted loss in the XC potential serves two purposes: it weighs the energetically relevant high-density region more and it makes both the loss terms have the same dimension (of energy). We model both $G^{\text{NNLDA}}$ and $G^{\text{NNGGA}}$ as feed-forward networks. The training data to model NNLLDA comprises of the exact density, XC energy, and XC potential for 5 atoms (Li, C, N, O, Ne) and two molecules (H$_2$ and LiH). The same training, albeit without the O atom, is used to model the NNGGA. 

\begin{figure}
    \centering
    \includegraphics{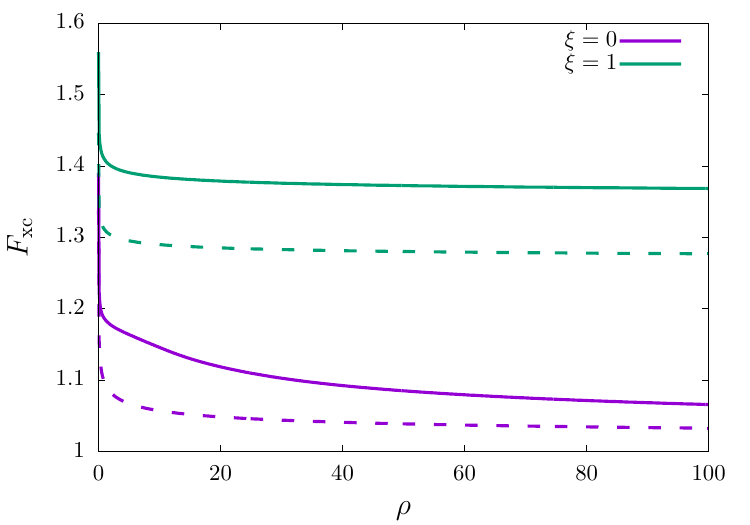}
    \caption{Comparison of enhancement factor of NNLDA (solid) with PW92 (dashed).}
    \label{fig:Fxc_NNLDA}
\end{figure}

\begin{figure}[tbhp!]
\centering
\begin{subfigure}{.5\textwidth}
  \centering
\includegraphics[scale=0.65]{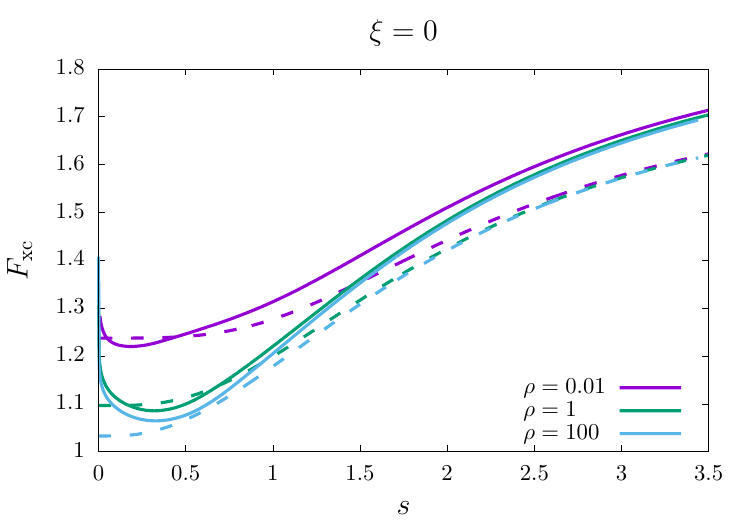}
  \label{fig:Fxc_NNGGA_xi_0}
\end{subfigure}%
\begin{subfigure}{.5\textwidth}
  \centering
    \includegraphics[scale=0.65]{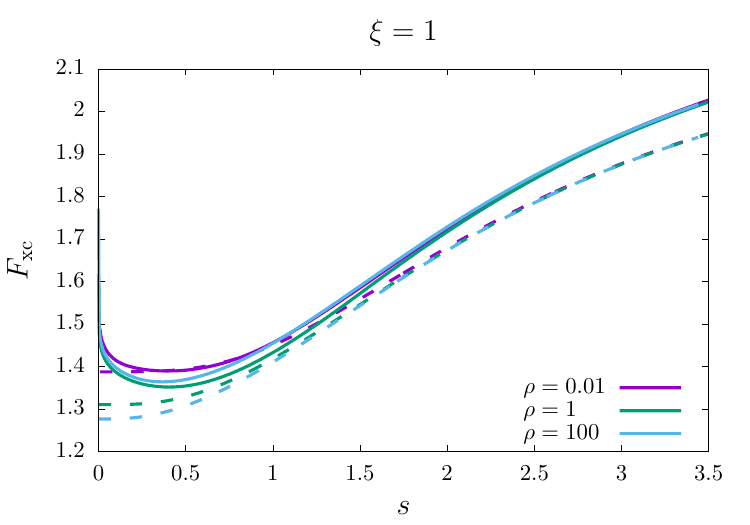}
  \label{fig:Fxc_NNGGA_xi_1}
\end{subfigure}
\caption{Comparison of enhancement factor of NNGGA (solid) with PBE (dashed).}
\label{fig:Fxc_NNGGA}
\end{figure}

\begin{figure}[tbhp!]
\centering
\begin{subfigure}{.5\textwidth}
  \centering
\includegraphics[scale=0.65]{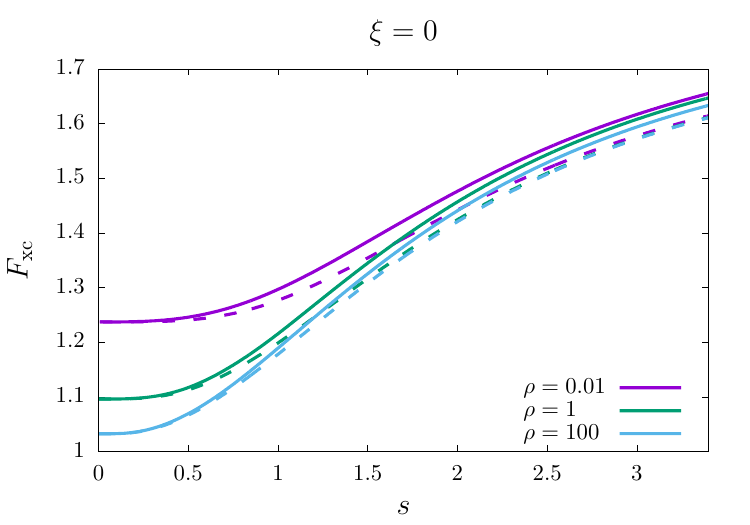}
  \label{fig:Fxc_NNGGA_UEG_xi_0}
\end{subfigure}%
\begin{subfigure}{.5\textwidth}
  \centering
    \includegraphics[scale=0.65]{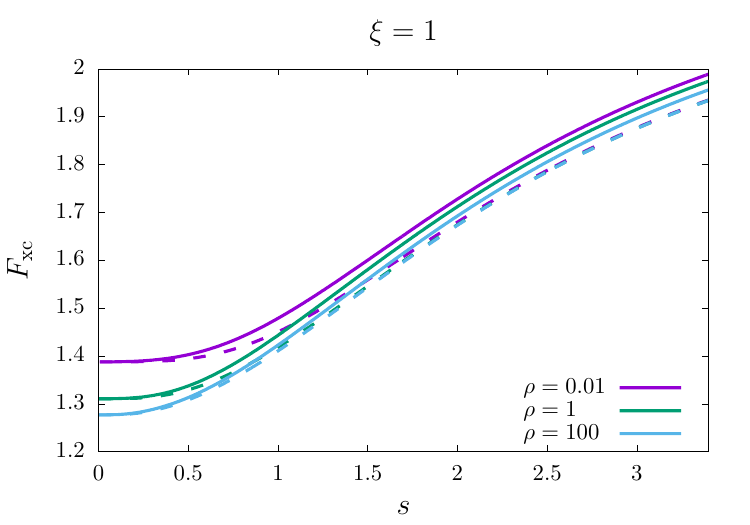}
  \label{fig:Fxc_NNGGA_UEG_xi_1}
\end{subfigure}
\caption{Comparison of enhancement factor of NNGGA-UEG (solid) with PBE (dashed).}
\label{fig:Fxc_NNGGA_UEG}
\end{figure}

\noindent Figures~\ref{fig:Fxc_NNLDA} and~\ref{fig:Fxc_NNGGA} present a comparison of the enhancement factor $F_{\text{xc}}[\rho](\br)=\exc[\rho](\br)/e_{\text{x}}^{\text{UEG}}[\rho](\br)$ for the resultant NNLDA and NNGGA functionals against the PW92 and PBE functionals, respectively. Both NNLDA and NNGGA exhibit significant differences over PW92 and PBE, respectively, reflecting the difference in their modeling---one fitted on energies and potentials of atoms and molecules and the other fitted to UEG and asymptotic conditions. We note that the NNGGA, being fitted to atoms and molecules, does not satisfy the UEG limit and exhibits a sharp gradient in the $s\rightarrow0$ limit. While this poses no difficulty for SCF calculations on atoms and molecules where the UEG limit is a rarity, it can lead to non-convergence in solids. We, thereby, propose another NN-based GGA functional, named NNGGA-UEG, that explicitly satisfies the UEG and is given as
\begin{equation} \label{eq:NNGGA_UEG}
    e_{\text{xc}}^{\text{NNGGA-UEG}}[\rho](\br) = e_{\text{xc}}^{\text{PBE}}[\rho](\br) + e_{\text{x}}^{\text{UEG}}[\rho](\br)\phi(\br)\text{tanh}\left(s^{1.5}\right)G^{\text{NNGGA-UEG}}(\rho(\br),\xi(\br),s(\br))\,.
\end{equation}
The NNGGA-UEG differs from NNGGA in two aspects: (i) additional $\text{tanh}(s^{1.5})$ factor, which explicitly enforces the correct UEG as encapsulated in PBE; and (ii) usage of UEG energy and potential as an additional training data on top of those used to model NNGGA. As evident from Fig.~\ref{fig:Fxc_NNGGA_UEG}, the NNGGA-UEG exhibits differences from PBE while smoothly approaching it in the UEG limit. 

\subsection{Accuracy}
\begin{figure}
    \centering
    \begin{subfigure}{\textwidth}
    \centering
    \includegraphics[scale=0.9]{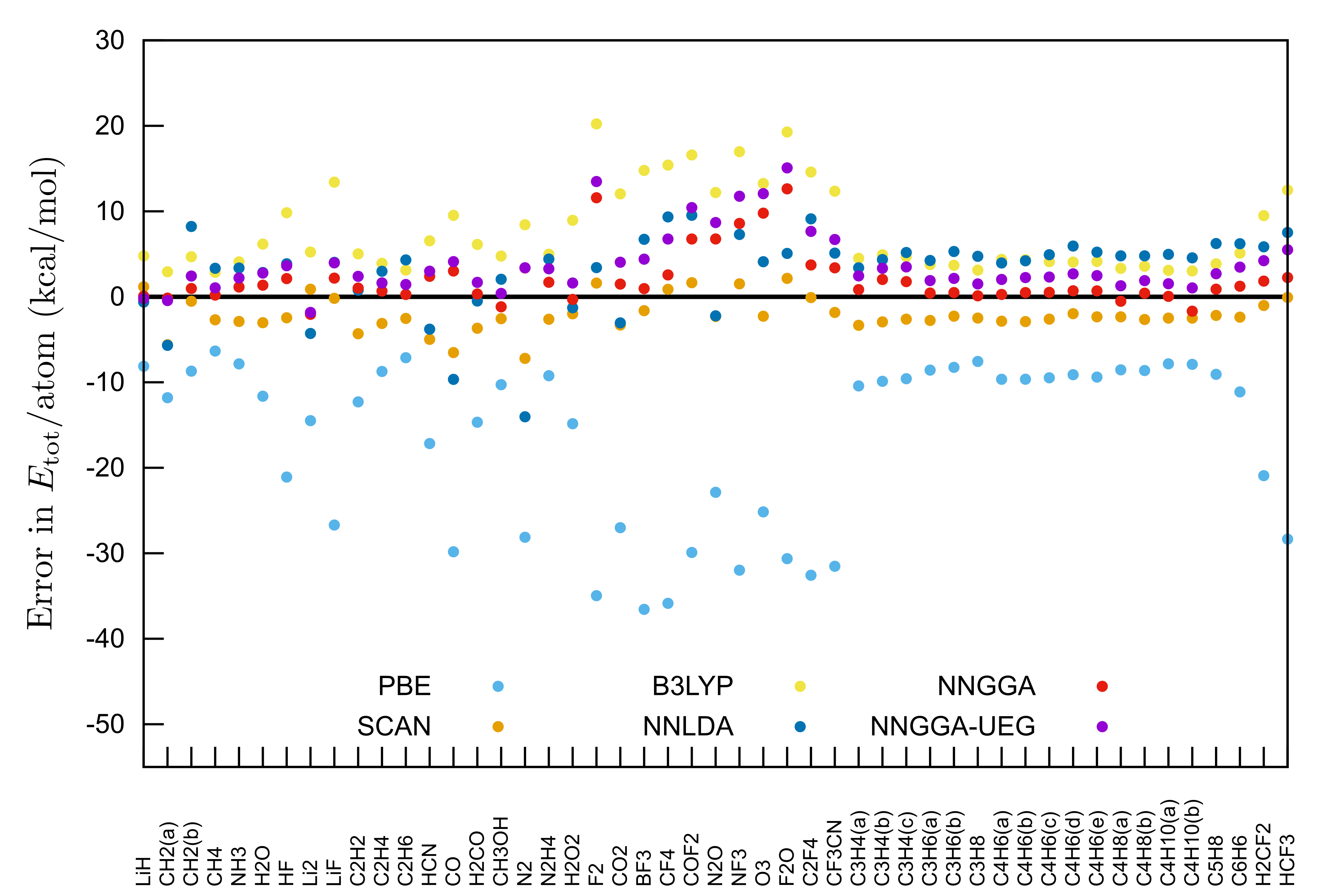}
    \caption{}
    \end{subfigure}
    \centering
    \begin{subfigure}{\textwidth}
    \includegraphics[scale=0.9]{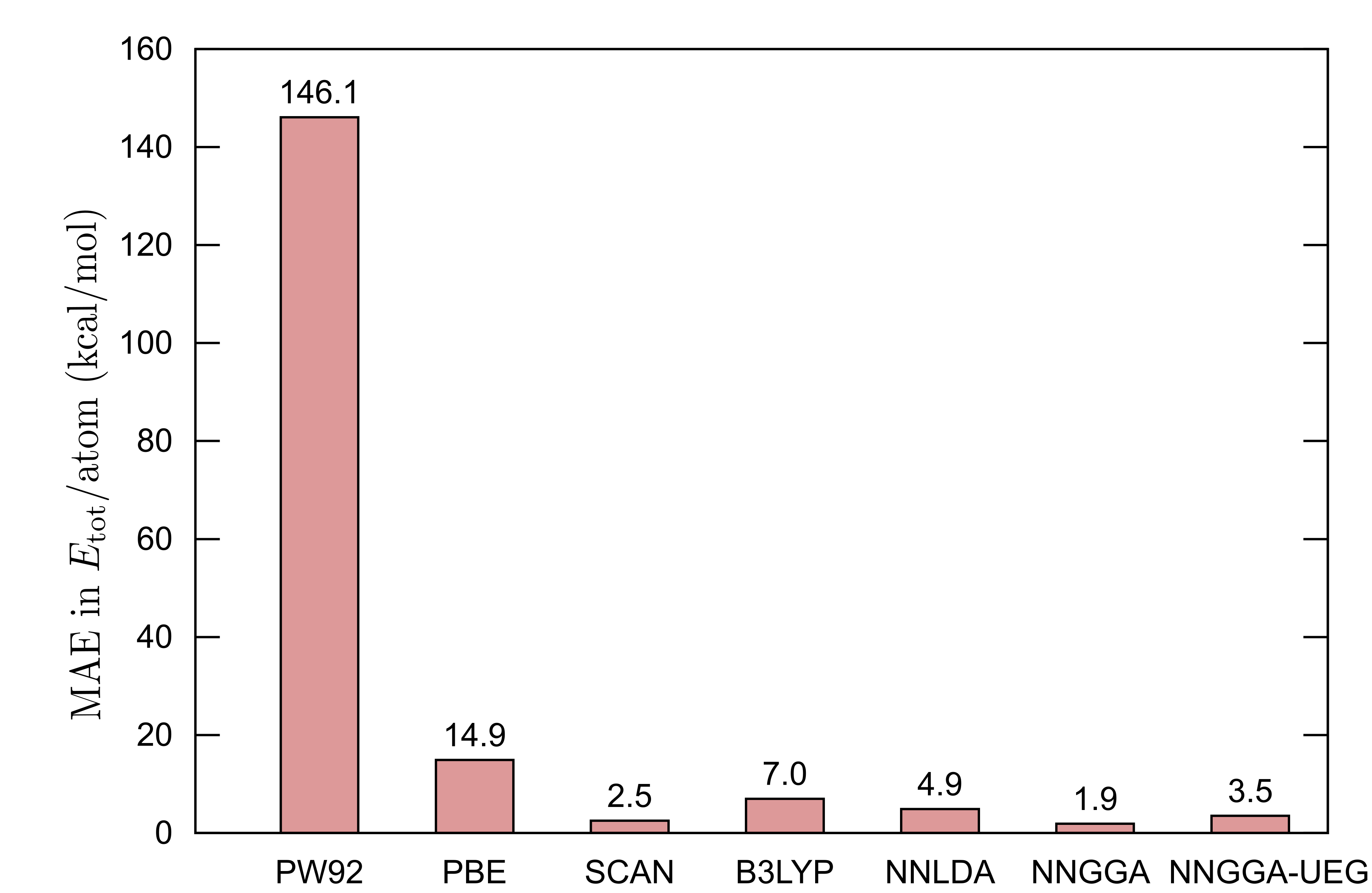} 
    \caption{}
    \end{subfigure}
    \caption{Comparison of error in total energy  ($E_{\text{tot}}$) per atom for molecules in the G2 dataset containing up to second-row elements (97 molecules). (a) Error in individual molecules. Molecules with a parenthesis indicate different isomers or spin-states. For clarity, we have shown the data for some of the molecules and have also omitted the comparison with PW92. See the SI for the remaining molecules, the comparison with PW92, and the details of individual systems. (b) Mean absolute error (MAE) for the 97 molecules in G2 containing up to second-row elements.}
    \label{fig:MAE_TE}
\end{figure}

\textbf{Total energies}: The NN-based functionals are assessed in terms of the total energies obtained at their self-consistently solved densities. This requires accurate groundstate energies from CI, which is difficult to attain beyond light atoms and some small molecules. However, accurate atomization energies are available via composite methods, such as G2~\cite{Curtiss1991}. This allows us to evaluate the total energies ($E_{\text{tot}}$) of molecules as $E_{\text{tot}} = \sum_A n_A E_A - \text{AE}$, where $E_A$ and $n_A$ are the energy and number of atoms of type $A$ in the molecule and $\text{AE}$ is the atomization energy. We obtain accurate values of $E_A$ from basis set extrapolation on CI energies for the atoms in the first two rows of the periodic table. Accurate $\text{AE}$ are obtained from the G2 dataset. This approach allows us to benchmark the accuracy of our NN-based functional using 97 out of 147 molecules in the G2 dataset, whose constituent atoms belong to the first two rows of the periodic table. As evident from Fig.~\ref{fig:MAE_TE}, the NN-based functionals attain impressive accuracy, exceeding even that of functionals in higher rungs (see the SI for details on individual molecules). The NNGGA, in particular, attains a mean absolute error (MAE) of 1.9 kcal/mol in $E_{\text{tot}}/\text{atom}$, surpassing both SCAN and B3LYP, and inching closer to the 1 kcal/mol mark of chemical accuracy. This remarkable generalization of the NN-based functionals trained on small training data speaks to the advantage of using the exact XC potential and energies. The significance of the greater accuracy of the NN-based functionals for total energy lies in the fact that any improvement in relative energies (e.g., atomization energy, barrier heights) will be on account of accurate energies and XC potentials for individual systems rather than relying on systematic cancellation of errors, as is the case with many XC functionals.   

\noindent \textbf{Quality of density}: We also assess the NN-based functionals based on the accuracy of their self-consistently solved density. While a widely used means of assessing the density is via the density-driven error as introduced in ~\cite{Kim2013}, it inherently includes a dependence on the functional itself, and hence, fails to delineate the purely density related errors. The mean-field error~\cite{Gould2023b}, instead, provides a more instructive measure of the quality of the densities defined solely in terms of the density. Given a density $\rho$, the mean-field energy is the non-XC part of the energy, given as $E_{\text{MF}}[\rho]=T_s[\rho]+\int \rho(\br)v_{\text{ext}}(\br)\dr + E_{\text{H}}[\rho]$. Correspondingly, given the exact density $\rho_{\text{exact}}$ and an approximate density $\rho_{\text{approx}}$, the mean-field error is $\Delta E_{\text{MF}}=E_{\text{MF}}[\rho_{\text{approx}}]-E_{\text{MF}}[\rho_{\text{exact}}]$. We note the presence of the non-interacting kinetic energy ($T_s[\rho]$) in $E_{\text{MF}}[\rho]$ requires an inverse DFT calculation to obtain the Kohn-Sham orbitals corresponding to $\rho$. As a result, an extensive study of the mean-field error for a wide range of molecules is difficult, given the limitations on getting $\rho_{\text{exact}}$ from CI and the high computational cost of inverse DFT. Nevertheless, we are able to perform inverse DFT on a few systems outside the training set of the NN-based functionals and obtain the mean-field error for the NN-based as well as other widely used functionals. As evident from Table~\ref{tab:MeanFieldErr}, for most systems, the NN-based functionals substantially improve upon their counterparts in the same rung. While SCAN provides the best accuracy, the NNGGA and NNGGA-UEG significantly improve upon PBE. Importantly, both NNGGA and NNGGA-UEG provide nearly the same accuracy as B3LYP. These results underscore that the NN-based models not just improve the energies but also the densities, a fact that can be attributed to the use of XC potentials in modeling them. 
\begin{table*} [htpb] 
 \begin{minipage}{\textwidth}
  \caption{Comparison of the quality of self-consistently solved density in terms of the  mean-field error.} 
  \label{tab:MeanFieldErr}
  \begin{tabular}{| M{2.5cm} | M{1.5cm} | M{1.5cm}| M{1.5cm} | M{1.5cm}| M{1.5cm} | M{1.5cm}| M{1.5cm}|}
    \hline
    System & PW92 & PBE & SCAN & B3LYP & NNLDA & NNGGA & NNGGA-UEG \\ \hline
    O & -42.88 & -25.85 & 1.01 & -6.17 & 11.93 & -8.76 & -8.35\\ \hline 
    F & -49.07 & -12.32 & 2.08 & -3.99 & 18.23 & -11.01 & -11.64\\ \hline 
    H$_2$(2eq)\footnote{H$_2$ molecule at 2$\times$ the equilibrium bond-length} & -10.42 & -5.31 & -5.23 & -5.99 & -6.34 & -4.75 & -4.56\\ \hline 
    BH & -14.11 & 3.58 & 8.36 & 6.77 & 1.38 & 5.48 & 5.68\\ \hline 
    H$_2$O & -23.94 & -5.62 & 2.35 & -1.7 & 11.23 & -6.13 & -4.06\\ \hline 
    CH$_2$ (triplet) & -24.87 & -5.14 & 1.3 & -5.31 & 0.81 & -2.56 & -2.85\\ \hline 
    CH$_2$ (singlet) & -19.48 & 0.1 & 5.32 & 2.94 & 1.1 & 1.76 & 2.27\\ \hline 
    CN & -40.26 & -8.96 & -0.44 & -9.36 & 65.13 & -6.19 & -5.81\\ \hline \hline 
    \textbf{MAE} & 28.13 & 8.36 & 3.26 & 5.28 & 14.52 & 5.83 & 5.65\\ \hline 
  \end{tabular}
   \end{minipage}
\end{table*}

\begin{figure}
    \centering
    \begin{subfigure}{\textwidth}
    \includegraphics{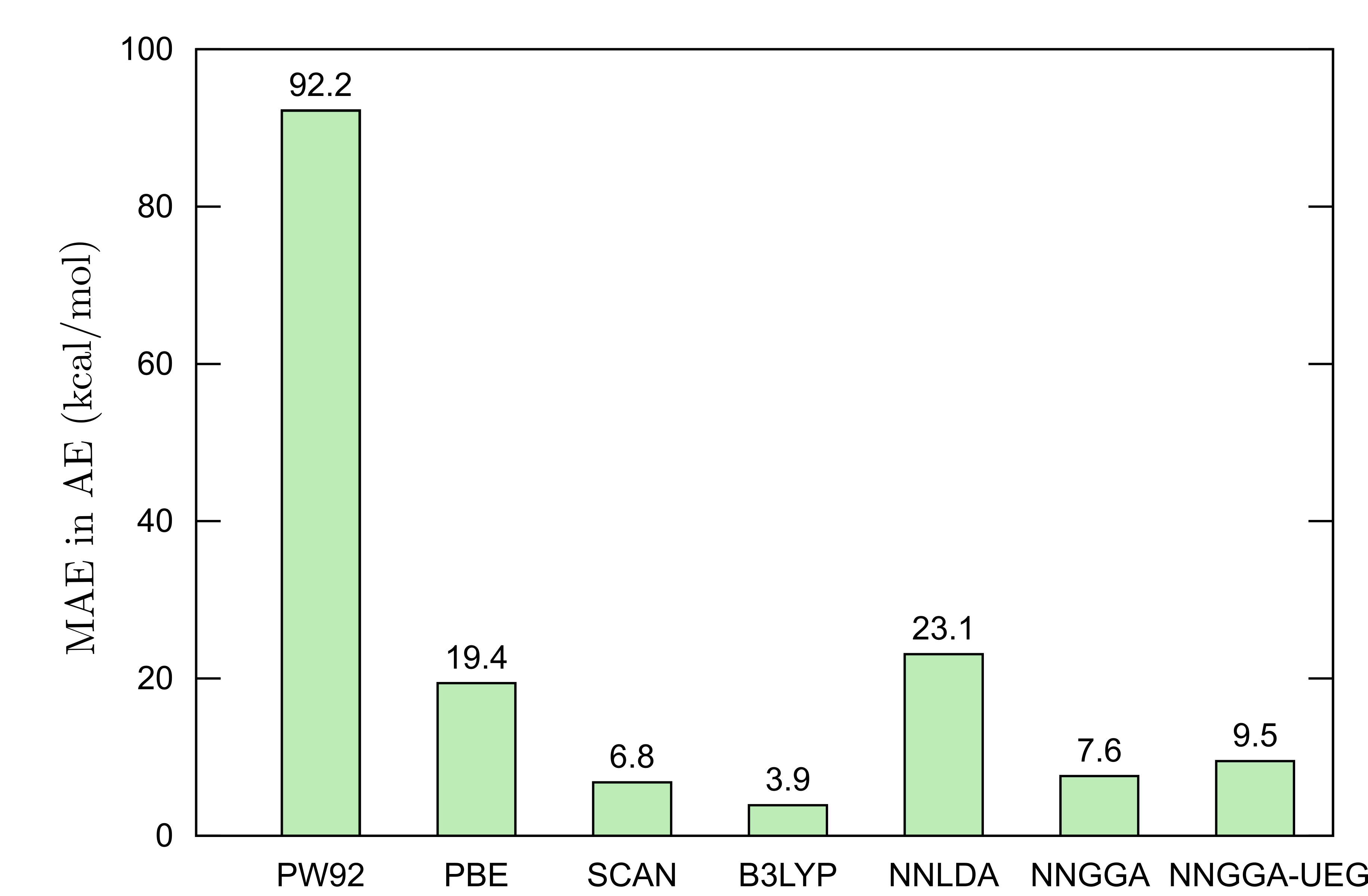}
    \caption{}
    \end{subfigure}
    \begin{subfigure}{\textwidth}
     \includegraphics{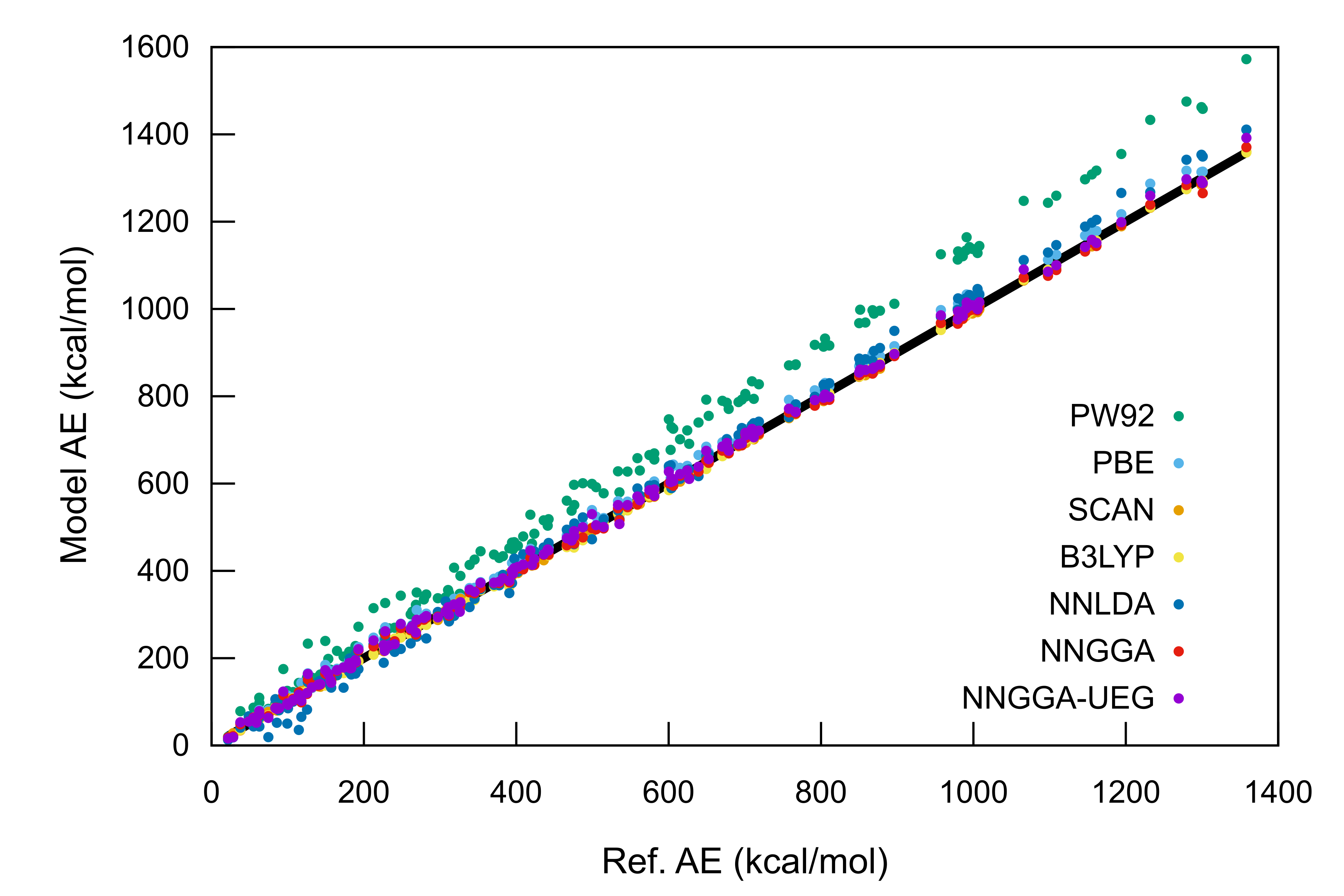}
    \end{subfigure}
    \caption{Comparison of the  atomization energy (AE) for the 147 molecules in the G2 dataset. (a) Mean absolute error (MAE). (b) Error in individual molecules. See the SI for details on each molecule.}
    \label{fig:MAE_AE}
\end{figure}

\begin{figure}
    \centering
    \includegraphics{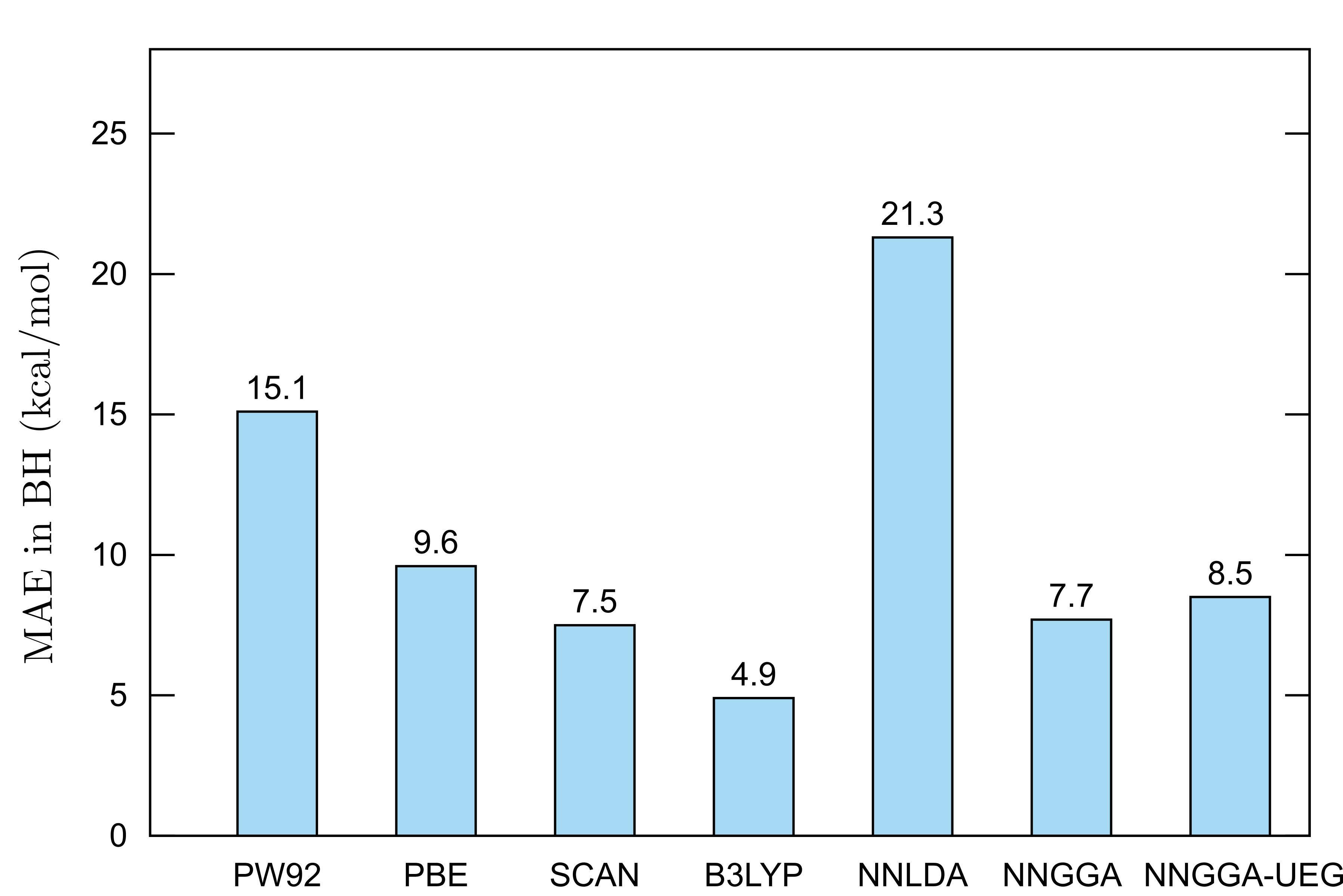}
    \caption{Comparison of the mean absolute error (MAE) in barrier height (BH) for the 60 neutral reactions in BH76 dataset.}
    \label{fig:MAE_BH}
\end{figure}

\noindent \textbf{Atomization energies and barrier heights}: While the NN-based functionals affords improvement in total energies and densities, it is important to measure their accuracy in relative energies, such as atomization energy and barrier heights, which govern much of chemistry. Fig.~\ref{fig:MAE_AE} compares the MAE in the atomization energy for the G2 dataset. All the NN-based models significantly improve over their counterparts in the same rung (NNLDA over PW92; NNGGA and NNGGA-UEG over PBE). In fact, the NN-based models provide almost the same accuracy as its next higher rung functional, i.e, NNLDA accuracy is close to PBE and NNGGA accuracy is close to SCAN. Fig.~\ref{fig:MAE_BH} compares the MAE in the barrier height for the 60 neutral reactions in BH76 dataset~\cite{Zhao2005}. We note that NNLDA, although better in total and atomization energies over PW92, performs worse for barrier heights. A plausible reasoning for it is that the NNLDA, being trained for atoms as well as molecules at equilibrium, generalizes well for the G2 dataset which primarily contain systems at equilibrium bond-length but not for BH76 wherein most of the transition states contain stretched bonds. The NNGGA and NNGGA-UEG, however, improve upon PBE, with NNGGA attaining the same accuracy as SCAN. We note that for both atomization energies and barrier heights the NNGGA-UEG offers slightly lower accuracy than NNGGA, highlighting the fact that at a GGA level simultaneously attaining good accuracy for thermochemistry while satisfying the UEG limit remains difficult. We refer to the SI for the details of the individual molecules and reactions considered in our atomization energy and barrier height benchmarks. 

\subsection{Discussion}
We have presented an accurate and transferable NN-based LDA and GGA functionals---namely NNLDA and NNGGA, NNGGA-UEG---trained using just five atoms and two molecules. In terms of total energy, quality of density, and atomization energy, the NN-based functionals significantly outperform widely used functionals in their respective rungs. Particularly, the NNGGA functional attains a remarkable accuracy of $<2$ kcal/mol in mean-absolute error (MAE) in total energy per atom, surpassing even that of higher rungs functionals like SCAN (meta-GGA) and B3LYP (hybrid). The improved accuracy in total energies signifies that any improvements for the NN-based functionals in relative energies will stem from attaining accuracy for individual systems rather than through systematic cancellation of errors. The NN-based functionals also improve over functionals in their rungs in terms of the quality of their self-consistently solved densities. In particular, the NNGGA and NNGGA-UEG functionals attain nearly the same accuracy as that of B3LYP in terms of the mean-field error in their densities, highlighting improvements not just in energies but also in the densities. Even for atomization energies and barrier heights, the NN-based GGA functionals substantially outperform the PBE functional, with the NN-based GGA functionals attaining nearly the same accuracy as the higher rung SCAN functional. These results underscore the promise of our approach of combining machine learning with the knowledge of exact densities, XC potentials and energies. Since the NN-based functionals were trained primarily on atoms and molecules, they might not attain similar gains for solids. In fact, at a GGA level, given the limited expressivity, it is well-known that there is a trade-off between doing well for thermochemistry (atomizatiion energy, barrier heights) and solid-state properties (lattice constant, cohesive energies)~\cite{Perdew2008}. We intend to address this by augmenting our current training data with those of solid-state systems in a future work. While the current work explores only up to the GGA-level, it offers a systematic way of constructing increasingly expressive and accurate functionals. In future studies, we seek to explore higher rungs (meta-GGA, non-local functionals) to find more accurate functionals that can also address the more severe challenges in DFT, such as self-interaction correction, accurate charge transfer, and bond dissociation. 
\section{Methods} \label{sec:methods}
\textbf{CI densities and energies}: We use heat bath configuration interaction~\cite{Chien2018,Dang2022,Dang2023}
(HBCI) to generate accurate densities of the atoms and molecules used in training as well as testing the mean-field error (see Table~\ref{tab:MeanFieldErr}). For all the systems, the HBCI calculations are done using the core-valence polarized, quadruple zeta Gaussian basis set (cc-pCVQZ)~\cite{Woon1995,peterson2002}. The HBCI begins from a reference state and systematically expands the wavefunction towards the full CI (FCI) limit. Although many implementations of HBCI start from a single-determinant reference, our implementation uses a complete active space (CAS) reference, up to a CAS (8e,8o) space.~\cite{roos1980,roos1980_2}. Our HBCI algorithm is implemented in a development version of the QChem software package~\cite{QChem5}. The details of the tolerances used in the HBCI calculations (e.g., for addition of new determinants and for measuring perturbative importance) are provided in the Supplemental Material in~\cite{Kanungo2023}. We obtain accurate ground state energies through the two-point Riemann zeta function based complete basis set extrapolation~\cite{Lesiuk2019}, using the energy values from cc-pCVTZ and cc-pCVQZ basis.    

\noindent \textbf{Inverse DFT}: In all our inverse DFT calculations, we employ spatially-adapted refined fourth-order spectral finite-elements to discretize the Kohn-Sham orbitals as well as their corresponding adjoint functions~\cite{Kanungo2019}. The XC potential, being much smoother in nature, is discretized using linear finite elements. At each iteration (i.e., guess for the XC potential), we solve the linear Kohn-Sham eigenvalue problem using the Chebyshev polynomial-based filtering technique~\cite{Zhou2006, MOTAMARRI2013308}. The adjoint problem, being a linear system of equations, is solved using the conjugate-gradient method. The update to the XC potential is obtained using the limited-memory Broyden-Fletcher-Goldfarb-Shanno (BFGS) algorithm~\cite{Nocedal1980}. The inverse DFT calculation is deemed converged when $||\rhoCI-\rhoKS||_{L_2} < 10^{-4}$.

\noindent \textbf{Neural network (NN) training}: We model the $G^{\text{NNLDA}}$,$G^{\text{NNGGA}}$, and $G^{\text{NNGGA-UEG}}$ using a feed-forward network of 3 hidden layers of 80 neurons with the exponential linear unit (ELU) activation function. The output layer is linear. The feed-forward networks are implemented via the Pytorch software package~\cite{Paszke2019}. The weights and biases are optimized using the Adam optimizer~\cite{Kingma2014} with a learning rate of $10^{-3}$. The weights and biases are deemed to be converged when the RMSE over 100 consecutive epochs reaches $10^{-3}$.

\noindent\textbf{Self-consistent field (SCF) calculations}: The SCF calculations involving the NN-based functionals are performed by porting the NN from Pytorch to  the DFT-FE software package~\cite{MOTAMARRI2020106853, das2022dft, Das2023}. We also perform the SCF calculations for PW92 and PBE using DFT-FE. For all the DFT-FE based calculations, we choose higher-order adaptive finite-element basis  which guarantees $<1$ kcal/mol in the ground state energy. The SCF calculations using SCAN and B3LYP XC functionals are performed using the QChem software package with the polarized quintuple zeta Gaussian basis (cc-pV5Z)~\cite{peterson2002}, so as to ensure $\sim$1 kcal/mol accuracy in the ground state energies.

\section{Data availability}
The files containing the various NN-based XC functionals are available at\\ \href{https://github.com/bikashkanungo/mlxc/tree/main}{https://github.com/bikashkanungo/mlxc/tree/main}. The geometries and the total energies of the various molecules considered in this study are provided as JSON files included in the Supplemental Information. The exact densities and XC potentials used as training of the NN-based XC functionals in this study are available upon reasonable request to the corresponding author.  

\section{Code Availability}
The code to perform the inverse DFT calculations is available upon reasonable request to the corresponding author. The demonstration code on how to use the NN-based XC functionals are provided at \href{https://github.com/bikashkanungo/mlxc/tree/main}{https://github.com/bikashkanungo/mlxc/tree/main}. 

\section*{Acknowledgements}
We acknowledge the support of Department of Energy, Office of Science, through grant number DE-SC0022241, under the auspices of which this study was conducted. B.K and V.G also acknowledge the support of Toyota Research Institute that supported the initial framework development used in this study. This study used resources of the NERSC Center, a DOE Office of Science User Facility supported by the Office of Science of the U.S. Department of Energy under Contract No. DE-AC02-05CH11231. This study also used resources of the Oak Ridge Leadership Computing Facility, which is a DOE Office of Science User Facility supported under Contract DE-AC05-00OR22725. 



\pagebreak
\begin{center}
\textbf{\Large Supplementary Information}
\end{center}

\setcounter{equation}{0}
\setcounter{figure}{0}
\setcounter{table}{0}
\setcounter{page}{1}
\setcounter{section}{0}
\makeatletter
\renewcommand{\theequation}{S\arabic{equation}}
\renewcommand{\thefigure}{S\arabic{figure}}
\renewcommand{\thetable}{S\arabic{table}}
\renewcommand{\thesection}{S\arabic{section}}
\renewcommand{\thepage}{S-\Roman{page}}

\section{Comparison of total energies}
\begin{figure}[htbp]
    \centering
    \begin{subfigure}{\textwidth}
    \centering
     \includegraphics[scale=1]{figs/g2_te_allowed_1.png}   
    \end{subfigure}
    \begin{subfigure}{\textwidth}
    \centering
    \includegraphics[scale=1]{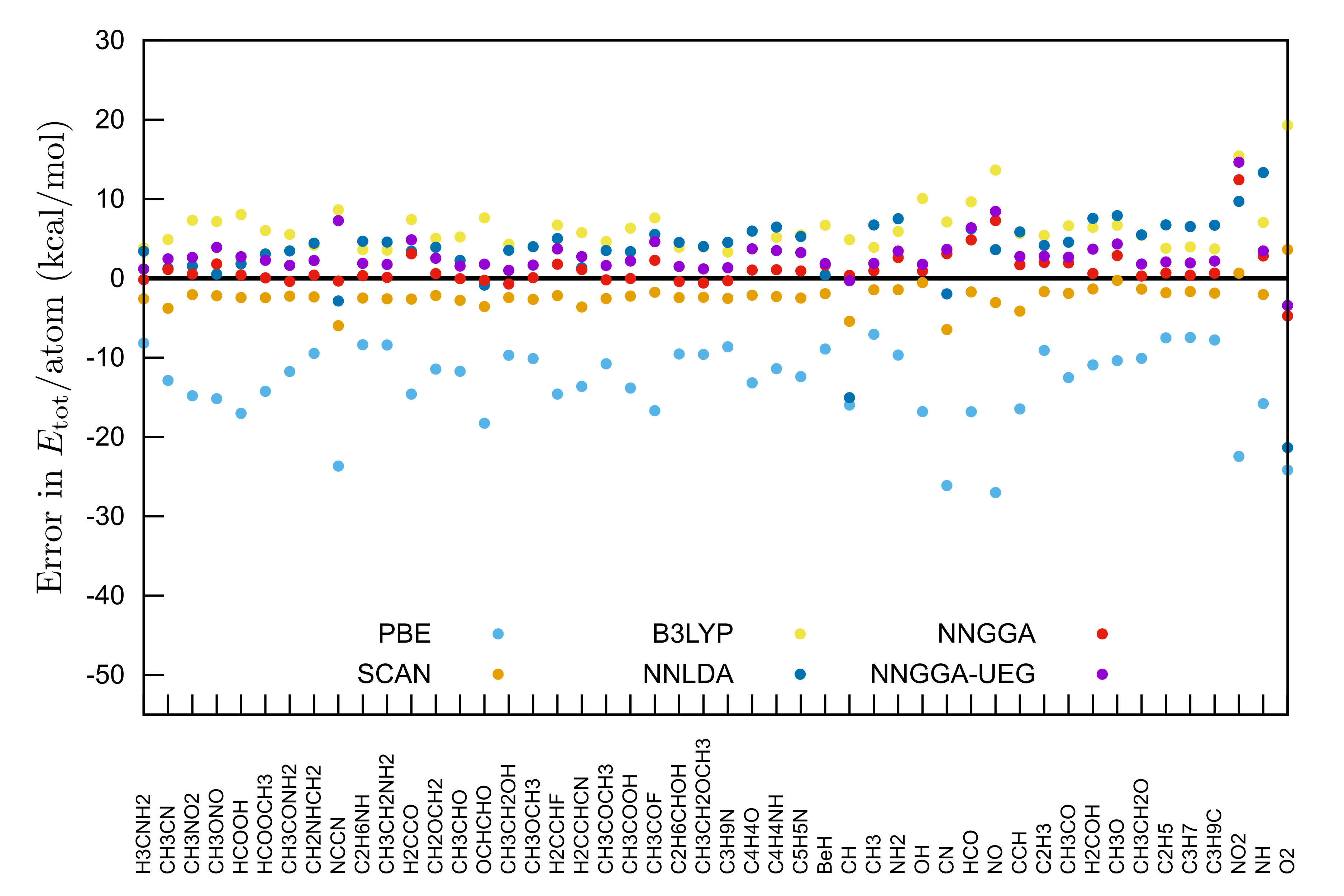}    
    \end{subfigure}
    \caption{Comparison of error in total energy  ($E_{\text{tot}}$) per atom for molecules in the G2 dataset containing up to second-row elements (97 molecules). For clarity we split the 97 molecules into two plots. Different isomers and spin-states of a molecule are distinguished with an added small-case letter inside a parenthesis.}
    \label{SI_fig:TE}
\end{figure}

\begin{figure}[htbp]
    \centering
    \begin{subfigure}{\textwidth}
    \centering
     \includegraphics[scale=1]{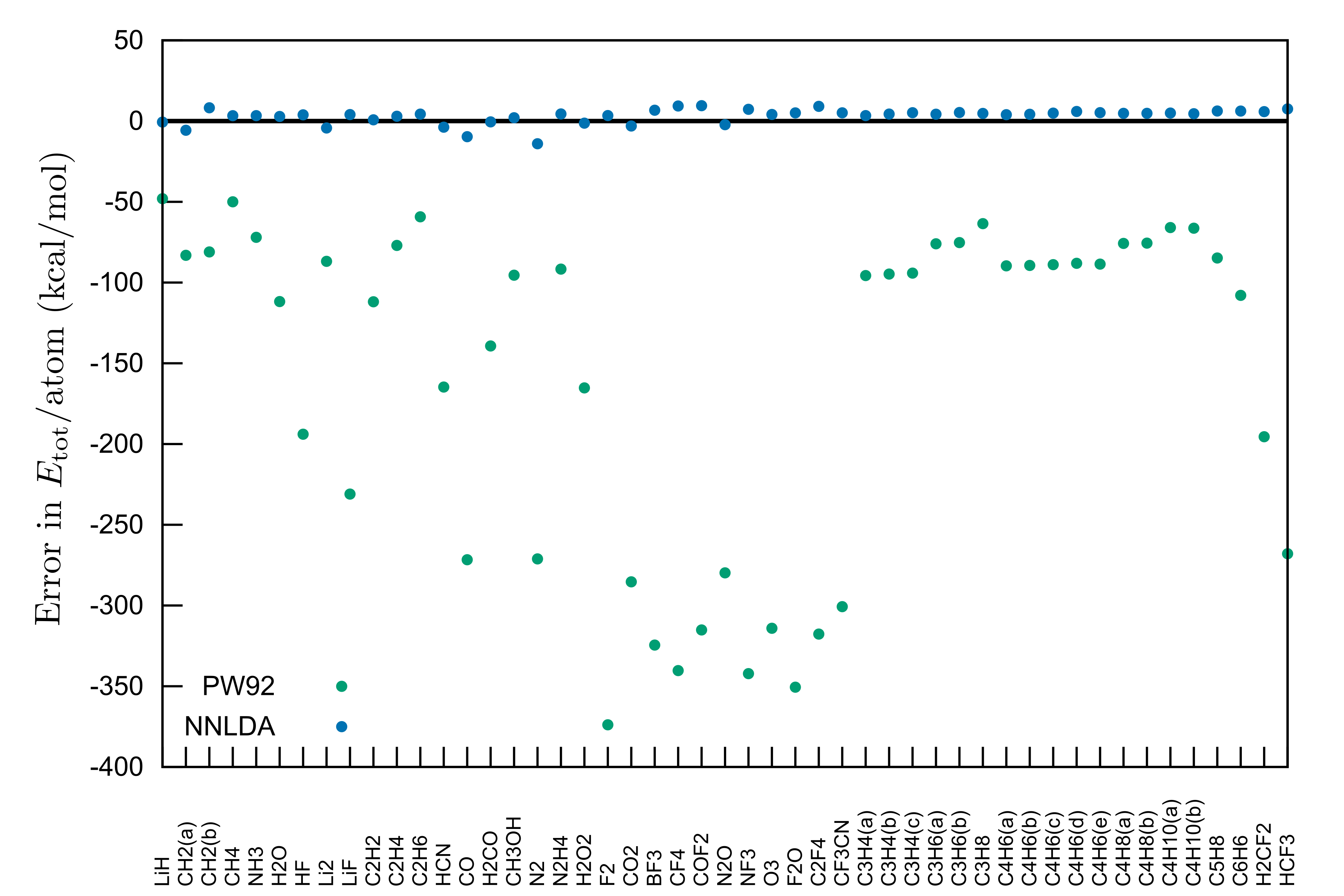}   
    \end{subfigure}
    \begin{subfigure}{\textwidth}
    \centering
    \includegraphics[scale=1]{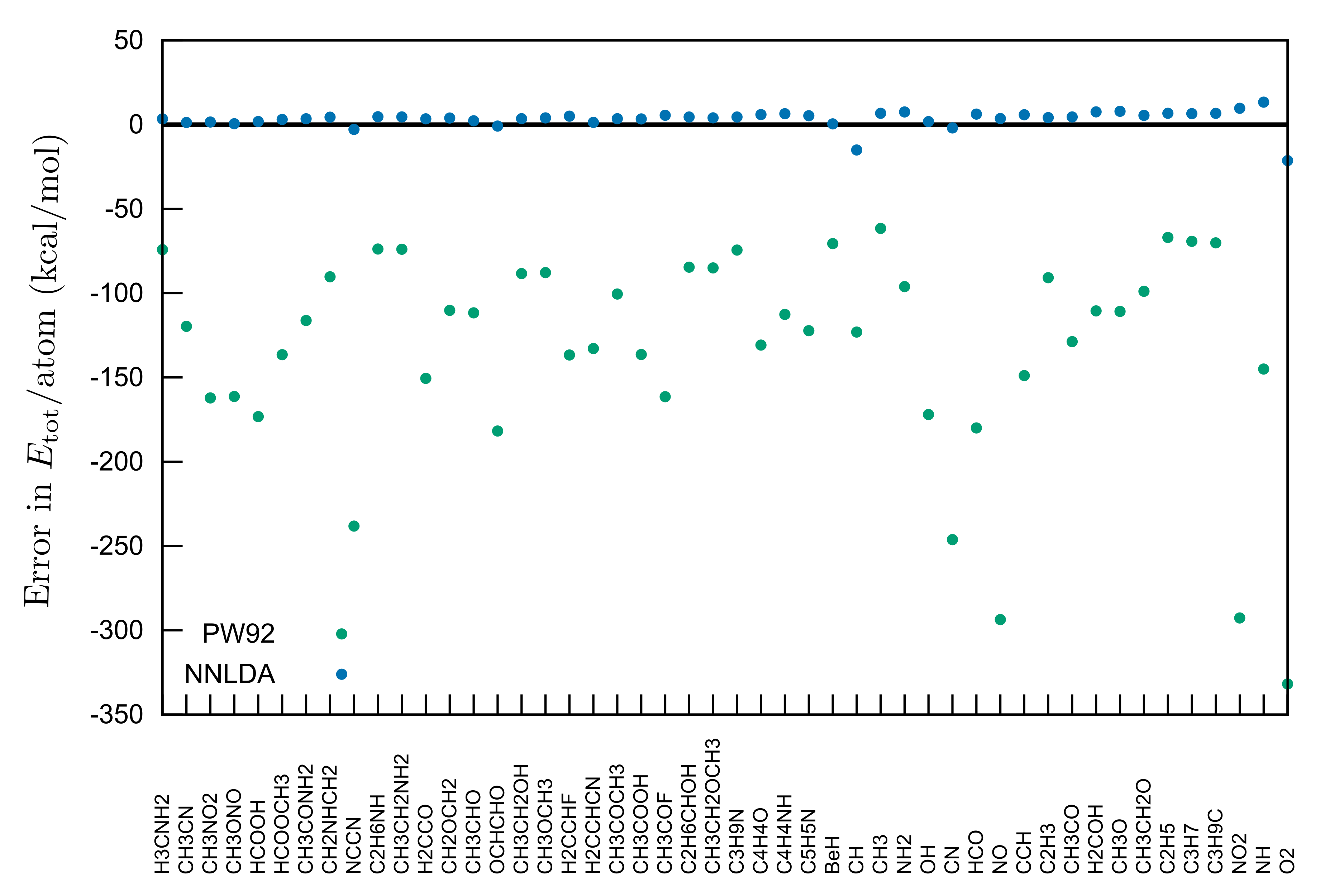}    
    \end{subfigure}
    \caption{Comparison between PW92 and NNLDA in their error in total energy  ($E_{\text{tot}}$) per atom for molecules in the G2 dataset containing up to second-row elements (97 molecules). For clarity we split the 97 molecules into two plots. Different isomers and spin-states of a molecule are distinguished with an added small-case letter inside a parenthesis.}
    \label{SI_fig:TE_PW92}
\end{figure}

Figure~\ref{SI_fig:TE} presents the error in total energies per atom for various XC functionals for the molecules in the G2 dataset containing up to second-row elements (97 out 147 molecules). For clarity, we present the comparison between PW92 and NNLDA separately in Fig.~\ref{SI_fig:TE_PW92}. In these figures, different isomers and spin-states of a molecule are distinguished with an added small-case letter inside a parenthesis. 

\section{Data for individual molecules}
The total energies, atomization energies, and barrier heights for individual molecules considered in this work are provided in $\texttt{TE_AE_BH.xlsx}$. Additionally, the geometries used for the G2 and BH76 molecules are provided in JSON format in the files named $\texttt{G2_geom.json}$ and $\texttt{BH76_geom.json}$, respectively. 

\end{document}